\documentclass{article}%
\usepackage{amsmath}
\usepackage{amsfonts}
\usepackage{amssymb}
\usepackage{graphicx}%
\begin{document}

\author{Sawa Manoff\\Bulgarian Academy of Sciences\\Institute for Nuclear Research and Nuclear Energy\\Department of Theoretical Physics\\Laboratory of Solitons, Coherency, and Geometry\\1784 Sofia - Bulgaria}
\date{E-mail address: smanov@inrne.bas.bg}
\title{Determination of the velocity of an emitter in spaces with affine connections
and metrics}
\maketitle
\tableofcontents

\begin{abstract}
Doppler effect and Hubble effect in different models of space-time related to
the space-time velocity of an observer are considered. The Doppler effect and
Doppler shift frequency parameter are connected with the kinematic
characteristics of the relative velocity of the emitter. The Hubble effect and
Hubble shift frequency parameter are considered in analogous way. It is shown
that by the use of the variation of the shift frequency parameter during a
time period, considered locally in the proper frame of reference of an
observer, one can directly determine the radial (centrifugal, centripetal)
relative velocity \ and acceleration as well as the tangential (Coriolis)
relative velocity and acceleration of an astronomical object moving relatively
to the observer. All results are obtained on purely kinematic basis without
taking into account the dynamic reasons for the considered effect.


\end{abstract}

\section{\bigskip Introduction}

1. Modern problems of relativistic astrophysics as well as of relativistic
physics (dark matter, dark energy, evolution of the universe, measurement of
velocities of moving objects etc.) are related to the propagation of signals
in space or in space-time \cite{Weinberg}, \cite{Unzicker}. The basis of
experimental data received as results of observations of the Doppler effect or
of the Hubble effect gives rise to theoretical considerations about the
theoretical status of effects related to detecting signals from emitters
moving relatively to observers carrying detectors in their laboratories.

2. All considerations related to the relative motions of objects with respect
to each other are made on the basis of the notions of relative velocity and
relative acceleration. The relative velocity is usually considered as radial
(centrifugal, centripetal) velocity and tangential (transversal, Coriolis)
velocity. The radial velocity is one of the most important concept in
astronomy and "a quantity whose precision of determination has improved
significantly in recent years. Its measuring is generally understood as the
object's motion along the line of sight, a quantity normally deduced from
observed spectral-line displacements, interpreted as Doppler shifts."
\cite{Lindegren}

The naive concept of radial velocity from point of view of different
theoretical models as the line-of-sight component of the stellar velocity
vector measured by the Doppler shift of the spectral lines is for too
simplistic when aiming at accuracies much better than $1$ $km.sec^{-1}$.

The need for a stringent definition of radial velocity leads to a definition
of this notion accepted by the $24^{th}$ General Assembly of the International
Astronomical Union in year $2000$ as a velocity of objects related to the s.c.
barycentric radial velocity measure, defined in a Barycentric Celestial
Reference System (BCRS) for accurate modelling of motions and events within
the solar system. This system (BCRS) of co-ordinates interpreted as frame of
reference serves as a quasi-Euclidean reference frame for the motions of
nearby stars and more distance objects. The radial velocity could be measured
by the use of two different methods:

(a) Spectroscopic method related to the Doppler effect. This method leads to
the notion of spectrometric radial velocity.

Since the recently generally accepted theories of gravitation have no adequate
formalism for description of the transversal Doppler effect, the notion of
tangential \ (transversal, Coriolis) velocity is not uniquely defined for
astronomical purposes. A spectroscopic line shift measurement, considered
until now, is equivalent to a direct comparison of the proper time scale at
the emitter and the observer. The reason for this is the existence of special
and general relativity theories considering the relations between different
co-ordinate and proper times. For more complicated accelerated motions between
observer and emitter (source) the relation between their proper time scales is
not uniquely determined because the notions related to their definitions are
defined, in general, as quantities not covariant or form invariant with resect
to the change of their frames of reference \cite{Lindegren}.

(b) Geometric method related to the change of lengths and angles between an
observer and moving objects. This method leads to the notions of kinematic
radial (centrifugal, centripetal) velocity and kinematic \ tangential
(transversal, Coriolis) velocity. The difficulties for the use of this method
appear in astronomy when the relative velocity of objects at very large
distances from Earth have to be measured.

The notion of astrometric radial velocity refers to the variation of the
co-ordinates of the source (emitter), and therefore, depends on the chosen
co-ordinate system and time scale. By contrast, the outcome of spectroscopic
observation is a directly measurable quantity and, therefore, independent of
co-ordinate systems \cite{Lindegren}.

3. Therefore, there are two different approaches for measuring radial
velocities of cosmic objects: a covariant approach related to proper times
(spectroscopic method) and a co-ordinate approach related to time and
distances in a co-ordinate system identified as the proper frame of reference
of the Sun (geometric method).

The covariant method is then specialized for the defined co-ordinate system.
This leads to difficulties related to the different definitions of
spectroscopic and kinematic quantities which are not directly connected to the
real measurement of the radial or tangential velocities of astronomical
objects. The comparison between the co-ordinate quantities and the
spectroscopic data leads to introduction of notions such as "barycentric
radial-velocity measure" and "astrometric radial velocity", where the notion
of real radial velocity is avoided. The reason for the introduction of the
above notions is the lack of covariant method for describing, on the one side,
the relative velocities and accelerations and, on the other side, the lack of
relations between these kinematic characteristics and their corresponding
Doppler shifts and Hubble shifts. Recently, it has been shown that the
introduced in $(\overline{L}_{n}$,$g)$-spaces kinematic characteristics
related to relative velocities and relative accelerations could be in simple
way expressed in terms of radial (centrifugal, centripetal) and tangential
(transversal, Coriolis) velocities and accelerations \cite{Manoff-1}. On the
other side, these kinematic characteristics could be related to an observer.
The question arises how could one observer find his own space-time velocity
and space-time acceleration with respect to an emitter (source) if the
spectral data corresponding to the Doppler shifts and Hubble shifts are
available. These velocity and acceleration could be considered as the velocity
and acceleration of the emitter from point of view of the observer. This could
be done because the observer, at first, determines the velocity and
acceleration of the emitter from his proper frame of reference on the basis of
the shifts of the signals of the emitter, measured in observer's frame of
reference, and then, on the basis of the relations between these shifts and
the motion of the observer in space-time, the observer could find his
space-time velocity and acceleration.

4. In the classical (non-quantum) field theories different models of
space-time have been used for describing the physical phenomena and their
evolution. The $3$-dimensional Euclidean space $E_{3}$ is the physical space
used as the space basis of classical mechanics \cite{Javorski}. The
$4$-dimensional (flat) Minkowskian space $\overline{M}_{4}$ is used as the
model of space-time in special relativity \cite{Tonnelat}. The (pseudo)
Riemannian spaces $V_{4}$ without torsion are considered as models of
space-time in general relativity \cite{Anderson}, \cite{Misner}. In
theoretical gravitational physics (pseudo) Riemannian spaces without torsion
as well as (pseudo) Riemannian spaces $U_{4}$ with torsion are proposed as
space-time grounds for new gravitational theories. To the most sophisticated
models of space-time belong the spaces with one affine connection and metrics
[$(L_{n},g)$-spaces] and the spaces with affine connections and metrics
[$(\overline{L}_{n},g)$-spaces].

The spaces with one affine connection and metrics [$(L_{n},g)$-spaces] have
affine connections whose components differ only by sign for contravariant and
covariant tensor fields over a differentiable manifold $M$ with $\dim~M=n$.

The spaces with affine connections and metrics have affine connections whose
components differ not only by sign for contravariant and covariant tensor
fields over a differentiable manifold $M$ with $\dim~M=n$.

3. Recently, it has been shown \cite{Manoff-1}, \cite{Manoff-2} that every
differentiable manifold $M$ ($dimM=n$) with affine connections and metrics
[$(\overline{L}_{n},g)$-spaces] \cite{Manoff-3} could be used as a model of
space-time for the following reasons:

\begin{itemize}
\item The equivalence principle (related to the vanishing of the components of
an affine connection at a point or on a curve) holds in $(\overline{L}_{n}%
,g)$-spaces \cite{Iliev-1}$\div$\cite{Iliev-1b}.

\item $(\overline{L}_{n},g)$-spaces have structures similar to these in
(pseudo) Riemannian spaces without torsion [$V_{n}$-spaces] allowing for
description of dynamic systems and the gravitational interaction
\cite{Manoff-2}.

\item Fermi-Walker transports and conformal transports exist in $(\overline
{L}_{n},g)$-spaces as generalizations of these types of transports in $V_{n}%
$-spaces \cite{Manoff-5}, \cite{Manoff-6}.

\item A Lorentz basis and a light cone could not be deformed in $(\overline
{L}_{n},g)$-spaces as it is the case in $V_{n}$-spaces.

\item All kinematic characteristics related to the notions of relative
velocity and of relative acceleration could be worked out in $(\overline
{L}_{n},g)$-spaces without changing their physical interpretations in $V_{n}%
$-spaces \cite{Manoff-2}, \cite{Manoff-7}$\div$\cite{Manoff-8a}.

\item $(\overline{L}_{n},g)$-spaces include all types of spaces with affine
connections and metrics used until now as models of space-time.
\end{itemize}

4. If a $(\overline{L}_{n},g)$-space \ could be used as a model of space or of
space-time the question arises how signals could propagate in a space-time
described by a $(\overline{L}_{n},g)$-space. The answer of this question has
been given in \cite{Manoff-8a}, \cite{Manoff-8}, and \cite{Manoff-9}. By that
the signals are described by means of null (isotropic) vector fields. A signal
could be defined as a periodical process transferred by an emitter and
received by an observer (detector) \cite{Manoff-8}. A wave front could be
considered as a signal characterized by its wave vector (null vector) as it is
the case in the geometrical optics in a $V_{n}$-space \cite{Stephani}. All
results are obtained on \textit{purely kinematic basis} (s. \cite{Manoff-8a},
\cite{Manoff-12}, \cite{Manoff-11}) without taking into account the dynamic
reasons for the considered effect.

On the basis of the general results in the previous papers we can draw a rough
scheme of the relations between the kinematic characteristics of the relative
velocity and relative acceleration on the one side, and the Doppler effect and
the Hubble effect on the other. Here the following abbreviations are used:

CM - classical mechanics

SRT - special relativity theory

GRT - general relativity theory

CRT - classical relativity theory \cite{Manoff-1}, \cite{Manoff-2}.

%

\begin{figure}
[ptb]
\begin{center}
\includegraphics[
natheight=25.197500cm,
natwidth=19.235800cm,
height=14.0298cm,
width=10.0034cm
]%
{C:/swp50/Docs/Sygnals.jpg}%
\end{center}
\end{figure}

\bigskip

5. The considerations of the Doppler effect and of the Hubble effect show that
the Doppler effect is derived in the physical theories (with exception of
general relativity) as a result of the relative motion of an observer and an
emitter, sending signals to the observer, from point of view of the proper
frame of reference of the observer and its relations to the proper frame of
reference of the emitter. On the other side, the Hubble effect could be
considered as a result of the Doppler effect and the Hubble law assumed to be
valid in the corresponding physical theory. In a rough scheme the relations
between Doppler effect, \ Hubble effect, and Hubble law could be represented
as follows:

\medskip%

\begin{tabular}
[c]{llllllll}
& {\small Relative motion} &  & {\small Doppler effect} &  & {\small Hubble
effect} &  & {\small Hubble law}\\
& {\small characterized} &  & {\small characterized}\textit{\ } &  &
{\small characterized} &  & {\small characterized}\\
& {\tiny \ }{\small by} &  & {\small as} &  & {\small as} &  & \\
{\small CM} & {\small constant } & $\Rightarrow$ & corollary & $\rightarrow$ &
corollary & $\Leftarrow$ & by definition\\
& {\small relative velocity} &  &  &  &  &  & \\
{\small SRT} & {\small constant } & $\Rightarrow$ & corollary & $\rightarrow$
& corollary & $\Leftarrow$ & by definition\\
& {\small relative velocity} &  &  &  &  &  & \\
{\small GRT} & {\small change of the} & $\Rightarrow$ & corollary &
$\rightarrow$ & corollary & $\Leftarrow$ & by definition\\
& {\small metrics of space-time} &  &  &  &  &  & of the metrics\\
{\small CRT} & {\small relative velocity and} & $\Rightarrow$ & corollary &
$\rightarrow$ & corollary & $\leftarrow$ & as corollary\\
& {\small relative acceleration} & $\Downarrow$ &  &  &  &  & $\Uparrow$\\
&  & $\Rightarrow$ & $\longrightarrow$ & $\rightarrow$ & $\longrightarrow$ &
$\rightarrow$ & $\Uparrow$%
\end{tabular}

6. Since the Doppler effect and the Hubble effect as kinematic effects could
be described by different theoretical schemes and models of space-time the
rich mathematical tools of the spaces with affine connections and metrics,
considered as models of space-time, are used for description of both the
effects. The aim has been to work out a theoretical model of the Doppler
effect and of the Hubble effect as corollaries only of the relative motion
between emitter and observer determined by the kinematic characteristics of
the relative velocity and the relative acceleration \ between emitter and
observer from point of view of the proper frame of reference of the observer.
For this task the $(n-1)+1$ formalism has been used related to the world line
of an observer and its corresponding $n-1$ dimensional sub space interpreted
as the observed space in the proper frame of the observer \cite{Manoff-12}.

7. Our task in the present paper is to investigate the influence of the
Doppler effect and of the Hubble effect on the frequency shift parameter. In
section 2 the Doppler effect and shift frequency parameter are considered. In
Section 3 the Hubble effect and shift frequency parameter are considered. It
is shown that by the use of the shift frequency parameter, considered locally
in the proper frame of reference of an observer \cite{Manoff-12}, we can
directly determine the radial (centrifugal, centripetal) relative velocity
\ and acceleration as well as the tangential (Coriolis) relative velocity and
acceleration. Section 4 comprises some concluding remarks. Most of the details
and derivations omitted in this paper could be found in \cite{Manoff-8} and in
\cite{Manoff-9}.

\subsection{Abbreviation and symbols}

\begin{itemize}
\item The vector field $u$ is the velocity vector field of an observer: $u\in
T(M)$, $\dim~M=n$, $n=4$.

\item The contravariant vector field $v_{z}=\mp l_{v_{z}}\cdot n_{\perp
}=H\cdot l_{\xi_{\perp}}\cdot n_{\perp}=H\cdot\xi_{\perp}$ is orthogonal to
$u$ and collinear to $\xi_{\perp}$. It is called radial (centrifugal,
centripetal) relative velocity.

\item The function $H=H(\tau)$ is called Hubble function.

\item The contravariant vector field $a_{z}=\mp l_{a_{z}}\cdot n_{\perp
}=\overline{q}\cdot l_{\xi_{\perp}}\cdot n_{\perp}=\overline{q}\cdot\xi
_{\perp}$ is orthogonal to $u$ and collinear to $\xi_{\perp}$. It is called
radial (centrifugal, centripetal) relative acceleration.

\item The function $\overline{q}=\overline{q}(\tau)$ is called acceleration
function (parameter).

\item The contravariant vector field $v_{\eta c}=\mp l_{v_{\eta c}}\cdot
m_{\perp}=\overline{H}_{c}\cdot l_{\xi_{\perp}}\cdot m_{\perp}=\overline
{H}_{c}\cdot\eta_{\perp}$ is orthogonal to $u$ and to $\xi_{\perp}$. It is
called tangential (Coriolis) relative velocity.

\item The function $\overline{H}_{c}=\overline{H}_{c}(\tau)$ is called
Coriolis Hubble function.{}

\item The contravariant vector field $a_{\eta_{c}}=\mp l_{a_{\eta c}}\cdot
m_{\perp}=\overline{q}_{\eta c}\cdot l_{\xi_{\perp}}\cdot m_{\perp}%
=\overline{q}_{\eta c}\cdot\eta_{\perp}$ is orthogonal to $u$ and to
$\xi_{\perp}$. It is called tangential (Coriolis) relative acceleration.

\item The function $\overline{q}_{\eta c}=\overline{q}_{\eta c}(\tau)$ is
called Coriolis acceleration function (parameter).

\item $\overline{\omega}$ is the frequency of a signal emitted by an emitter
at a time $\tau-d\tau$ of the proper time of the observer.

\item $\omega$ is the frequency of a signal detected by the observer at a time
$\tau$ of the proper time of the observer.

\item $d\tau$ is the time interval in the proper frame of reference of the
observer for which the signal propagates from the emitter to the observer
(detector) at a space distance $dl$ in the proper frame of reference of the observer.

\item $\tau$ is the proper time parameter of the observer in his proper frame
of reference.
\end{itemize}

\section{Doppler effect and shift frequency parameter}

It has been shown \cite{Manoff-8}, \cite{Manoff-9} that in a $(\overline
{L}_{n},g)$-space longitudinal (standard) and transversal Doppler effects
could appear when signals are propagating from an emitter to an observer
(detector) moving relatively to each other.

\subsubsection{Longitudinal (standard) Doppler effect and the shift frequency
parameter}

1. Let a signal with frequency $\overline{\omega}$ be emitted by an emitter
\cite{Manoff-8} at the time $\tau-d\tau$ and be received by an observer
(detector) at the time $\tau$.

Since $\overline{\omega}=\omega(\tau-d\tau)$ and $\omega=\omega(\tau)$, we can
expand $\overline{\omega}$ in a Taylor row up to the second order of $d\tau$%
\[
\overline{\omega}=\omega(\tau-d\tau)\approx\omega(\tau)-\frac{d\omega}{d\tau
}_{\mid\tau}\cdot d\tau+\frac{1}{2}\cdot\frac{d^{2}\omega}{d\tau^{2}}%
_{\mid\tau}\cdot d\tau^{2}+O(d\tau)\text{ \ \ .}%
\]

Then%
\[
d\omega=\overline{\omega}-\omega\approx-\frac{d\omega}{d\tau}\cdot d\tau
+\frac{1}{2}\cdot\frac{d^{2}\omega}{d\tau^{2}}\cdot d\tau^{2}\text{ \ \ ,}
\]%
\[
\frac{d\omega}{\omega}=\frac{\overline{\omega}-\omega}{\omega}\approx-\frac
{1}{\omega}\cdot\frac{d\omega}{d\tau}\cdot d\tau+\frac{1}{2}\cdot\frac
{1}{\omega}\cdot\frac{d^{2}\omega}{d\tau^{2}}\cdot d\tau^{2}\text{ \ \ \ \ .}
\]

On the other side, the results in the case of a general motion of the observer
read \cite{Manoff-8}, \cite{Manoff-9}%
\begin{align*}
\frac{\overline{\omega}-\omega}{\omega} &  =\frac{d\omega}{\omega}=\\
&  =-\frac{dl}{l_{u}}\cdot(\frac{l_{v_{z}}}{l_{\xi_{\perp}}}+\frac
{l_{(a_{\perp})_{z}}}{l_{u}})+\frac{1}{2}\cdot\frac{dl^{2}}{l_{u}^{2}}%
\cdot(\frac{l_{a_{z}}}{l_{\xi_{\perp}}}+\frac{l_{(\nabla_{u}a)_{\perp z}}%
}{l_{u}})\text{ \ \ \ , }\\
\text{\ \ \ \ }dl &  =l_{u}\cdot d\tau\text{ \ \ \ \ .}%
\end{align*}

Since%
\[
\frac{dl}{l_{u}}=d\tau\text{ \ \ \ \ , \ \ \ \ \ \ \ \ }\frac{dl^{2}}%
{l_{u}^{2}}=d\tau^{2}\text{ \ \ \ ,}%
\]
we obtain the relation%
\[
\overline{\omega}=\omega\cdot\lbrack1-(\frac{l_{v_{z}}}{l_{\xi_{\perp}}}%
+\frac{l_{(a_{\perp})_{z}}}{l_{u}})\cdot d\tau+\frac{1}{2}\cdot(\frac
{l_{a_{z}}}{l_{\xi_{\perp}}}+\frac{l_{(\nabla_{u}a)_{\perp z}}}{l_{u}})\cdot
d\tau^{2})]\text{ \ \ \ \ \ .}%
\]

2. If we consider only infinitesimal changes of the frequency $\omega$ for the
time interval $d\tau$, i.e. if $d\omega=\overline{\omega}-\omega$, we can
express the shift parameter $z=(\overline{\omega}-\omega)/\omega$ as an
infinitesimal quantity%
\begin{align*}
z  &  =\frac{d\omega}{\omega}=d(\log~\omega)=\\
&  =d\overline{z}=-\frac{1}{\omega}\cdot\frac{d\omega}{d\tau}\cdot d\tau
+\frac{1}{2}\cdot\frac{1}{\omega}\cdot\frac{d^{2}\omega}{d\tau^{2}}\cdot
d\tau^{2}=\\
&  =-(\frac{l_{v_{z}}}{l_{\xi_{\perp}}}+\frac{l_{(a_{\perp})_{z}}}{l_{u}%
})\cdot d\tau+\frac{1}{2}\cdot(\frac{l_{a_{z}}}{l_{\xi_{\perp}}}%
+\frac{l_{(\nabla_{u}a)_{\perp z}}}{l_{u}})\cdot d\tau^{2}\text{ \ \ .}%
\end{align*}

On the other side, $d\overline{z}$ could be considered as a differential of
the function $\overline{z}$ depending on the proper time $\tau$ of the
observer, i.e. $\overline{z}=\overline{z}(\tau)$. It is assumed that
$\overline{z}(\tau)$ has the necessary differentiability properties. The
function $\overline{z}$ at the point $\overline{z}(\tau-d\tau)$ could be
represented in Taylor row as
\[
\overline{z}(\tau-d\tau)=\overline{z}(\tau)-\frac{d\overline{z}}{d\tau}\cdot
d\tau+\frac{1}{2}\cdot\frac{d^{2}\overline{z}}{d\tau^{2}}\cdot d\tau
^{2}+O(d\tau)\text{ \ \ .\ \ }%
\]
Then $\overline{z}(\tau-d\tau)$ and $d\overline{z}=\overline{z}(\tau
-d\tau)-\overline{z}(\tau)$ could be written up to the second order of $d\tau$
respectively as%

\begin{align*}
\overline{z}(\tau-d\tau)  &  =\overline{z}(\tau)-\frac{d\overline{z}}{d\tau
}\cdot d\tau+\frac{1}{2}\cdot\frac{d^{2}\overline{z}}{d\tau^{2}}\cdot
d\tau^{2}\text{ \ \ \ ,}\\
d\overline{z}  &  =\overline{z}(\tau-d\tau)-\overline{z}(\tau)=\\
&  =-\frac{d\overline{z}}{d\tau}\cdot d\tau+\frac{1}{2}\cdot\frac
{d^{2}\overline{z}}{d\tau^{2}}\cdot d\tau^{2}=\\
&  =-\frac{1}{\omega}\cdot\frac{d\omega}{d\tau}\cdot d\tau+\frac{1}{2}%
\cdot\frac{1}{\omega}\cdot\frac{d^{2}\omega}{d\tau^{2}}\cdot d\tau^{2}=\\
&  =-(\frac{l_{v_{z}}}{l_{\xi_{\perp}}}+\frac{l_{(a_{\perp})_{z}}}{l_{u}%
})\cdot d\tau+\frac{1}{2}\cdot(\frac{l_{a_{z}}}{l_{\xi_{\perp}}}%
+\frac{l_{(\nabla_{u}a)_{\perp z}}}{l_{u}})\cdot d\tau^{2}\text{ \ \ .}%
\end{align*}

The comparison of the coefficients before $d\tau$ and $d\tau^{2}$ in the last
(above) two expressions leads to the relations%
\[
\frac{d\overline{z}}{d\tau}=\frac{1}{\omega}\cdot\frac{d\omega}{d\tau}\text{
\ \ \ \ , \ \ \ \ \ \ }\frac{d^{2}\overline{z}}{d\tau^{2}}=\frac{1}{\omega
}\cdot\frac{d^{2}\omega}{d\tau^{2}}\text{ \ \ \ ,\ }%
\]%
\[
\frac{d\overline{z}}{d\tau}=\frac{l_{v_{z}}}{l_{\xi_{\perp}}}+\frac
{l_{(a_{\perp})_{z}}}{l_{u}}\text{ \ \ \ \ , \ \ \ \ \ \ \ \ \ \ }\frac
{d^{2}\overline{z}}{d\tau^{2}}=\frac{l_{a_{z}}}{l_{\xi_{\perp}}}%
+\frac{l_{(\nabla_{u}a)_{\perp z}}}{l_{u}}\text{\ \ .}%
\]

The vector $\xi_{\perp}$ could be chosen as a unit vector, i.e. $l_{\xi
_{\perp}}=1$, equal to the vector $n_{\perp}$ showing the direction to the
emitter from point of view of the observer. Then%
\[
\frac{d\overline{z}}{d\tau}=l_{v_{z}}+\frac{l_{(a_{\perp})_{z}}}{l_{u}}\text{
\ \ \ \ \ \ , \ \ \ \ \ \ \ \ \ \ }\frac{d^{2}\overline{z}}{d\tau^{2}%
}=l_{a_{z}}+\frac{l_{(\nabla_{u}a)_{\perp z}}}{l_{u}}\text{ \ \ \ \ \ .\ }%
\]

Therefore, if we can measure the change (variation) of the shift frequency
parameter $d\overline{z}$ in a time interval $d\tau$ we can find the
centrifugal (centripetal) relative velocity and radial (centrifugal,
centripetal) relative acceleration of the emitter with respect to the
observer. The above relations appear as direct way for a check-up of the
considered theoretical scheme of the propagation of signals in spaces with
affine connections and metrics. On the other side, the explicit form of
$l_{v_{z}}$ and $l_{a_{z}}$ as functions of the kinematic characteristics of
the relative velocity and relative acceleration could lead to conclusions of
the properties of the space-time model used for description of the physical
phenomena. The same relations could lead to more precise assessment of the
Hubble function $H$ and the acceleration function $\overline{q}$ at a given time.

3. Let us now consider the shift frequency parameter when the observer's world
line is an auto-parallel trajectory, i.e. when the velocity vector $u$ of the
observer fulfills the equation%
\[
\nabla_{u}u=f\cdot.u\text{ \ \ \ \ \thinspace, \ \ \ \ \ \ }f\in C^{\infty
}(M)\text{ .}%
\]

Then, because of $\overline{g}[h_{u}(u)]=0$,%
\[
a_{\perp}=\overline{g}[h_{u}(a)]=f\cdot\overline{g}[h_{u}(u)]=0\text{ \ \ \ ,}%
\]%
\begin{align*}
(\nabla_{u}a)_{\perp}  &  =\overline{g}[h_{u}(\nabla_{u}(f\cdot u))]=\overline
{g}[h_{u}((uf)\cdot u+f\cdot a)]=\\
&  =(uf)\cdot\overline{g}[h_{u}(u)]+f\cdot\overline{g}[h_{u}(a)]=0\text{
\ \ \ \ \ .}%
\end{align*}

In this case, the relations are fulfilled%
\[
\frac{d\overline{z}}{d\tau}=\frac{l_{v_{z}}}{l_{\xi_{\perp}}}\text{ \ \ \ \ ,
\ \ \ \ \ \ \ \ \ \ }\frac{d^{2}\overline{z}}{d\tau^{2}}=\frac{l_{a_{z}}%
}{l_{\xi_{\perp}}}\text{\ \ .}%
\]%
\[
\frac{d\overline{z}}{d\tau}=l_{v_{z}}\text{ \ \ \ \ \ \ , \ \ \ \ }\frac
{d^{2}\overline{z}}{d\tau^{2}}=l_{a_{z}}\text{ \ , \ \ \ }l_{\xi_{\perp}%
}=1\text{ \ . }%
\]

Therefore, if we can measure the change (variation) of the shift frequency
parameter $d\overline{z}_{c}$ in a time interval $d\tau$ we can find the
radial (centrifugal, centripetal) relative velocity and the radial
(centrifugal, centripetal) relative acceleration of the emitter with respect
to the observer moving on an auto-parallel world line.

\subsubsection{Transversal Doppler effect and the shift frequency parameter}

In analogous way we can find the relations between the absolute values of the
tangential (Coriolis) relative velocity and the tangential (Coriolis) relative
acceleration and the shift frequency parameter.

1. The relation between the frequency $\overline{\omega}$ of the emitted
signals and the frequency $\omega$ of the detected signals reads
\cite{Manoff-8}, \cite{Manoff-9}%
\begin{align*}
\overline{\omega}  &  =\omega\cdot\lbrack1-\frac{dl}{l_{u}}\cdot(\frac
{1}{l_{\xi_{\perp}}}\cdot l_{v_{\eta c}}+\frac{1}{l_{u}}\cdot l_{(a_{\perp
})_{c}})+\frac{1}{2}\cdot\frac{dl^{2}}{l_{u}^{2}}\cdot(\frac{l_{a_{\eta c}}%
}{l_{\xi_{\perp}}}+\frac{1}{l_{u}}\cdot l_{(\nabla_{u}a)_{\perp c}})]=\\
&  =\omega\cdot\lbrack1-(\frac{l_{v_{\eta c}}}{l_{\xi_{\perp}}}+\frac{1}%
{l_{u}}\cdot l_{(a_{\perp})_{c}})\cdot d\tau+\frac{1}{2}\cdot(\frac{l_{a_{\eta
c}}}{l_{\xi_{\perp}}}+\frac{1}{l_{u}}\cdot l_{(\nabla_{u}a)_{\perp c}})\cdot
d\tau^{2}]\text{ \ \ \ .}%
\end{align*}

The shift frequency parameter has the form%
\begin{align*}
z_{c}  &  =\frac{\overline{\omega}-\omega}{\omega}=-(\frac{l_{v_{\eta c}}%
}{l_{\xi_{\perp}}}+\frac{1}{l_{u}}\cdot l_{(a_{\perp})_{c}})\cdot d\tau
+\frac{1}{2}\cdot(\frac{l_{a_{\eta c}}}{l_{\xi_{\perp}}}+\frac{1}{l_{u}}\cdot
l_{(\nabla_{u}a)_{\perp c}})\cdot d\tau^{2}=\\
&  =d\overline{z}_{c}=\\
&  =-\frac{d\overline{z}_{c}}{d\tau}\cdot d\tau+\frac{1}{2}\cdot\frac
{d^{2}\overline{z}_{c}}{d\tau^{2}}\cdot d\tau^{2}+O(d\tau)\approx\\
&  \approx-\frac{d\overline{z}_{c}}{d\tau}\cdot d\tau+\frac{1}{2}\cdot
\frac{d^{2}\overline{z}_{c}}{d\tau^{2}}\cdot d\tau^{2}\text{\ .}%
\end{align*}

The comparison of the coefficients before $d\tau$ and $d\tau^{2}$ in the two
expressions%
\begin{align*}
z_{c}  &  =\frac{\overline{\omega}-\omega}{\omega}=-(\frac{l_{v_{\eta c}}%
}{l_{\xi_{\perp}}}+\frac{1}{l_{u}}\cdot l_{(a_{\perp})_{c}})\cdot d\tau
+\frac{1}{2}\cdot(\frac{l_{a_{\eta c}}}{l_{\xi_{\perp}}}+\frac{1}{l_{u}}\cdot
l_{(\nabla_{u}a)_{\perp c}})\cdot d\tau^{2}=\\
&  =d\overline{z}_{c}=\\
&  \approx-\frac{d\overline{z}_{c}}{d\tau}\cdot d\tau+\frac{1}{2}\cdot
\frac{d^{2}\overline{z}_{c}}{d\tau^{2}}\cdot d\tau^{2}\text{ \ \ }%
\end{align*}
leads to the relations%
\[
\frac{d\overline{z}_{c}}{d\tau}=\frac{l_{v_{\eta c}}}{l_{\xi_{\perp}}}%
+\frac{1}{l_{u}}\cdot l_{(a_{\perp})_{c}}\text{ \ \ \ \ , \ \ \ \ \ }%
\frac{d^{2}\overline{z}_{c}}{d\tau^{2}}=\frac{l_{a_{\eta c}}}{l_{\xi_{\perp}}%
}+\frac{1}{l_{u}}\cdot l_{(\nabla_{u}a)_{\perp c}}\text{ \ \ .\ \ \ \ }%
\]

The vector $\xi_{\perp}$ could be chosen as a unit vector, i.e. $l_{\xi
_{\perp}}=1$, equal to the vector $n_{\perp}$ showing the direction to the
emitter from point of view of the observer. Then%
\[
\frac{d\overline{z}_{c}}{d\tau}=l_{v_{\eta c}}+\frac{1}{l_{u}}\cdot
l_{(a_{\perp})_{c}}\text{ \ \ \ \ , \ \ \ \ \ }\frac{d^{2}\overline{z}_{c}%
}{d\tau^{2}}=l_{a_{\eta c}}+\frac{1}{l_{u}}\cdot l_{(\nabla_{u}a)_{\perp c}%
}\text{ \ \ .\ }%
\]

Therefore, if we can measure the change (variation) of the shift frequency
parameter $d\overline{z}_{c}$ in a time interval $d\tau$ we can find the
tangential \  (Coriolis) relative velocity and the tangential (Coriolis)
relative acceleration of the emitter with respect to the observer. The above
relations appear as direct way for a check-up of the considered theoretical
scheme of the propagation of signals in spaces with affine connections and
metrics. On the other side, the explicit form of $l_{v_{\eta c}}$ and
$l_{a_{\eta c}}$ as functions of the kinematic characteristics of the relative
velocity and relative acceleration could lead to conclusions of the properties
of the space-time model used for description of the physical phenomena. The
same relations could lead to more precise assessment of the Hubble function
$H_{c}$ and the acceleration function $\overline{q}_{\eta c}$ at a given time.

2. In the case when the observer's world line is an auto-parallel trajectory,
the relations are fulfilled%
\[
\frac{d\overline{z}_{c}}{d\tau}=\frac{l_{v_{\eta c}}}{l_{\xi_{\perp}}}\text{
\ \ \ \ , \ \ \ \ \ }\frac{d^{2}\overline{z}_{c}}{d\tau^{2}}=\frac{l_{a_{\eta
c}}}{l_{\xi_{\perp}}}\text{\ \ .}%
\]%
\[
\text{\ \ }\frac{d\overline{z}_{c}}{d\tau}=l_{v_{\eta c}}\text{ \ \ \ \ ,
\ \ \ \ \ }\frac{d^{2}\overline{z}_{c}}{d\tau^{2}}=l_{a_{\eta c}}\text{
\ ,\ \ \ \ \ \ }l_{\xi_{\perp}}=1\text{ \ . }%
\]

Therefore, if we can measure the change (variation) of the shift frequency
parameter $d\overline{z}_{c}$ in a time interval $d\tau$ we can find the
tangential (Coriolis) relative velocity and the tangential (Coriolis) relative
acceleration of the emitter with respect to the observer moving on an
auto-parallel world line.

\section{Hubble effect and shift frequency parameter}

It has been shown \cite{Manoff-8}, \cite{Manoff-9} that in a $(\overline
{L}_{n},g)$-space longitudinal (standard) and transversal Hubble effects could
appear when signals are propagating from an emitter to an observer (detector)
moving relatively to each other.

\subsection{Longitudinal (standard) Hubble effect and the shift frequency
parameter}

1. By the use of the relations \cite{Manoff-8}, \cite{Manoff-9}
\[
l_{v_{z}}=\mp H\cdot l_{\xi_{\perp}}\text{ \ \ \ \ \ \ , \ \ \ \ \ \ \ \ \ \ }%
l_{a_{z}}=\mp\overline{q}\cdot l_{\xi_{\perp}}\text{ \ \ \ \ \ ,}%
\]%
\begin{align*}
\frac{d\overline{z}}{d\tau}  &  =\frac{l_{v_{z}}}{l_{\xi_{\perp}}}%
+\frac{l_{(a_{\perp})_{z}}}{l_{u}}\text{ \ \ \ \ , \ \ \ \ \ \ \ \ \ \ }%
\frac{d^{2}\overline{z}}{d\tau^{2}}=\frac{l_{a_{z}}}{l_{\xi_{\perp}}}%
+\frac{l_{(\nabla_{u}a)_{\perp z}}}{l_{u}}\text{ \ \ \ \ \ \ ,}\\
l_{(a_{\perp})_{z}}  &  =g(a_{\perp},n_{\perp})\text{ \ \ \ \ \ \ \ \ ,
\ \ \ \ \ \ \ \ \ \ \ }l_{(\nabla_{u}a)_{\perp z}}=g(n_{\perp},(\nabla
_{u}a)_{\perp})\text{ \ \ \ ,}%
\end{align*}
we can find the Hubble function $H$ and the acceleration function (parameter)
$\overline{q}$ respectively as%
\begin{align*}
\frac{d\overline{z}}{d\tau}  &  =\mp H\text{ }+\frac{l_{(a_{\perp})_{z}}%
}{l_{u}}\text{\ \ \ \ , \ \ \ \ \ \ \ }\frac{d^{2}\overline{z}}{d\tau^{2}}%
=\mp\overline{q}+\frac{l_{(\nabla_{u}a)_{\perp z}}}{l_{u}}\text{ \ \ \ \ ,}\\
\frac{d\overline{z}}{d\tau}  &  =\mp H\text{ }+\frac{g(a_{\perp},n_{\perp}%
)}{l_{u}}=\mp\widetilde{H}\text{ \ \ \ \ \ , \ \ \ \ \ }\frac{d^{2}%
\overline{z}}{d\tau^{2}}=\mp\overline{q}+\frac{g(n_{\perp},(\nabla
_{u}a)_{\perp})}{l_{u}}=\mp\widetilde{q}\text{ \ \ ,}\\
\mp\widetilde{H}  &  =\mp H\text{ }+\frac{g(a_{\perp},n_{\perp})}{l_{u}}\text{
\ \ \ \ \ \ \ \ , \ \ \ \ \ \ \ \ \ }\mp\widetilde{q}=\mp\overline{q}%
+\frac{g(n_{\perp},(\nabla_{u}a)_{\perp})}{l_{u}}\text{ \ \ .}%
\end{align*}

2. If the world line of the observer is an auto-parallel trajectory then we
can find a direct relation between the change of the shift frequency parameter
and the Hubble function $H$ as well as the acceleration parameter
$\overline{q}$%
\[
\frac{d\overline{z}}{d\tau}=\mp H\text{ \ \ \ \ , \ \ \ \ \ \ \ }\frac
{d^{2}\overline{z}}{d\tau^{2}}=\mp\overline{q}\text{ \ \ .}%
\]

\subsection{Transversal Hubble effect and the shift frequency parameter}

1. By the use of the relations \cite{Manoff-8}, \cite{Manoff-9}%
\[
l_{v_{\eta c}}=\mp\overline{H}_{c}\cdot l_{\xi_{\perp}}\text{ \ \ \ \ \ ,
\ \ \ \ \ \ \ \ \ \ \ }l_{a_{\eta c}}=\mp\overline{q}_{\eta c}\cdot
l_{\xi_{\perp}}\text{ \ \ \ ,}%
\]%
\begin{align*}
\frac{d\overline{z}_{c}}{d\tau}  &  =\frac{l_{v_{\eta c}}}{l_{\xi_{\perp}}%
}+\frac{1}{l_{u}}\cdot l_{(a_{\perp})_{c}}\text{ \ \ \ \ , \ \ \ \ \ }%
\frac{d^{2}\overline{z}_{c}}{d\tau^{2}}=\text{\ }\frac{d^{2}\overline{z}_{c}%
}{d\tau^{2}}=l_{a_{\eta c}}+\frac{1}{l_{u}}\cdot l_{(\nabla_{u}a)_{\perp c}%
}\text{ \ \ \ ,}\\
l_{(a_{\perp})_{c}}  &  =g(a_{\perp},m_{\perp})\text{
\ \ \ \ \ \ \ \ \ \ \ \ \ \ \ , \ \ \ \ \ \ }l_{(\nabla_{u}a)_{\perp c}%
}=g(m_{\perp},(\nabla_{u}a)_{\perp})~\text{\ \ \ \ \ \ ,}%
\end{align*}
we can find the transversal Hubble function $H_{c}$ and the transversal
acceleration function (parameter) $\overline{q}_{\eta c}$ respectively as%
\begin{align*}
\frac{d\overline{z}_{c}}{d\tau}  &  =\mp\overline{H}_{c}\text{ }+\frac
{1}{l_{u}}\cdot l_{(a_{\perp})_{c}}\text{\ \ \ \ \ \ , \ \ \ \ \ \ \ \ \ }%
\frac{d^{2}\overline{z}_{c}}{d\tau^{2}}=\mp\overline{q}_{\eta c}+\frac
{1}{l_{u}}\cdot l_{(\nabla_{u}a)_{\perp c}}\text{ \ \ ,}\\
\frac{d\overline{z}_{c}}{d\tau}  &  =\mp\overline{H}_{c}\text{ }+\frac
{1}{l_{u}}\cdot g(a_{\perp},m_{\perp})=\mp\widetilde{H}_{c}\text{ \ \ \ \ ,
\ \ }\\
\text{\ \ \ \ }\frac{d^{2}\overline{z}_{c}}{d\tau^{2}}  &  =\mp\overline
{q}_{\eta c}+\frac{1}{l_{u}}\cdot g(m_{\perp},(\nabla_{u}a)_{\perp}%
)=\mp\widetilde{q}_{\eta c}\text{ \ ,}\\
\mp\widetilde{H}_{c}  &  =\mp\overline{H}_{c}\text{ }+\frac{1}{l_{u}}\cdot
g(a_{\perp},m_{\perp})\text{ \ \ , \ \ \ \ \ \ }\mp\widetilde{q}_{\eta c}%
=\mp\overline{q}_{\eta c}+\frac{1}{l_{u}}\cdot g(m_{\perp},(\nabla
_{u}a)_{\perp})\text{ \ ,\ \ \ \ }%
\end{align*}
\bigskip

Therefore, we have fine $4$ equations for the vector field $u$ as space-time
velocity of the observer if the first and second derivative of $\overline{z}$
and $\overline{z}_{c}$ are found by the experiments.
\begin{align*}
\frac{d\overline{z}}{d\tau}  &  =\mp H\text{ }+\frac{g(a_{\perp},n_{\perp}%
)}{l_{u}}\text{ \ \ ,}\\
\frac{d^{2}\overline{z}}{d\tau^{2}}  &  =\mp\overline{q}+\frac{g(n_{\perp
},(\nabla_{u}a)_{\perp})}{l_{u}}\text{ \ \ ,}\\
\frac{d\overline{z}_{c}}{d\tau}  &  =\mp\overline{H}_{c}\text{ }+\frac
{1}{l_{u}}\cdot g(a_{\perp},m_{\perp})\text{ \ ,}\\
\frac{d^{2}\overline{z}_{c}}{d\tau^{2}}  &  =\mp\overline{q}_{\eta c}+\frac
{1}{l_{u}}\cdot g(m_{\perp},(\nabla_{u}a)_{\perp})\text{ \ \ ,}%
\end{align*}
\bigskip where%

\begin{align*}
H  &  =\frac{1}{n-1}\cdot\theta\mp\sigma(n_{\perp},n_{\perp})\text{ \ \ ,}\\
\overline{H}_{c}  &  =\frac{1}{n-1}\cdot\theta\mp\sigma(m_{\perp},m_{\perp
})\text{ \ \ ,}\\
\overline{q}  &  =\frac{1}{n-1}\cdot U\mp~_{s}D(n_{\perp},n_{\perp})\text{
,}\\
\overline{q}_{\eta c}  &  =\mp~_{s}D(m_{\perp},m_{\perp})\text{ \ \ \ ,}\\
\pm l_{u}\cdot\frac{dl_{u}}{d\tau}  &  =g(u,a)+\frac{1}{2}\cdot(\nabla
_{u}g)(u,u)\text{ \ .}%
\end{align*}

2. If the world line of the observer is an auto-parallel trajectory then we
can find a direct relation between the change of the shift frequency parameter
and the Hubble function $\overline{H}_{c}$ as well as the acceleration
parameter $\overline{q}_{\eta c}$%
\[
\frac{d\overline{z}_{c}}{d\tau}=\mp\overline{H}_{c}\text{ \ \ \ \ \ ,
\ \ \ \ \ }\frac{d^{2}\overline{z}_{c}}{d\tau^{2}}=\mp\overline{q}_{\eta
c}\text{ \ \ .}%
\]

3. The finding out of $u$ by the use of the $4$-equations is based on the fact
that the quantities $H$, $H_{c}$, $\overline{q}$, $\overline{q}_{\eta c}$,
$g(n_{\perp},a_{\perp})$, $g(n_{\perp},(\nabla_{u}a)_{\perp})$, $g(a_{\perp
},m_{\perp})$, and $g(m_{\perp},(\nabla_{u}a)_{\perp})$ are functions of the
vector field $u$ and its covariant derivatives. If the affine connections and
metrics are given in a $(\overline{L}_{n},g)$-space then the solving of the
above equations could determine the space-time velocity of the observer with
respect to the emitter.

\section{Conclusion}

In the present paper we have considered the Doppler effect and the Hubble
effect in a $(\overline{L}_{n},g)$-space and their relations to the shift
frequency parameters corresponding to the longitudinal and transversal
effects. It is shown that these effects lead to direct check-up of the
theoretical scheme and could be used for finding out the relative radial
(centrifugal, centripetal) velocities and accelerations as well as the
relative tangential (transversal, Coriolis) velocities and accelerations of
moving astronomical objects from point of view of the proper frame of
reference of an observer (detector).

The Doppler effects and the Hubble effects are considered on the grounds of
purely kinematic considerations. It should be stressed that the Hubble
functions $H$ and $\overline{H}_{c}$ are introduced on a purely kinematic
basis related to the notions of relative radial (centrifugal, centripetal)
velocity and to the notions of tangential (Coriolis) velocities respectively.
They could be found directly by the use of the measurements of the shift
frequency parameters. It should be noted that $\overline{H}_{c}$ does not
exists in the Einstein theory of gravitation. The dynamic interpretations of
$H$ and $\overline{H}_{c}$ in a theory of gravitation depend on the structures
of the theory and the relations between the field equations and on both the
functions. In this paper, it is shown that notions the specialists use to
apply in theories of gravitation and cosmological models could have a good
kinematic grounds independent of any concrete classical field theory. Doppler
effects, and Hubble effects could be used in mechanics of continuous media and
in other classical field theories in the same way as the standard Doppler
effect is used in classical and special relativistic mechanics.

\end{document}